\documentclass[twocolumn,prl,showpacs]{revtex4}
\usepackage{graphicx}

\begin{document}
\title{Landau-Zener Tunnelling in Waveguide Arrays}

\author{Ramaz Khomeriki} 
\email{khomeriki@hotmail.com}
\affiliation{Department of Physics, Tbilisi State University, 3 Chavchavadze 
avenue, 
Tbilisi 0128, Republic of Georgia}

\author{Stefano Ruffo} 
\email{ruffo@avanzi.de.unifi.it}
\affiliation{Dipartimento di Energetica ``S. Stecco" and CSDC, 
Universit\'a di Firenze, and INFN, Via S. Marta, 3, 50139 Firenze, Italy}

\date{\today}

\begin{abstract}

Landau-Zener tunnelling is discussed in connection with optical waveguide arrays.
Light injected in a specific band of the Bloch spectrum in the propagation constant
can be transmitted to another band, changing its physical properties.
This is achieved using two waveguide arrays with different refractive indices,
which amounts to consider a Schr\"odinger equation in a periodic potential
with a step. The step causes wave ``acceleration" and thus induces Landau-Zener 
tunnelling. The region of physical parameters where this phenomenon can
occur is analytically determined and a realistic experimental setup is suggested. 
Its application could allow the realization of light filters.

\pacs{42.82.Et; 42.25.-p; 03.65.-w}

\end{abstract}

\maketitle

When a quantum system is subject to an external force, a
non-adiabatic crossing of energy levels can occur. This phenomenon is
known as {\it Landau-Zener tunnelling}~\cite{landau,zener} and some of its 
recent observations are for Josephson junctions~\cite{kieran} and optical 
effective two-level systems~\cite{opt}.
On the other hand, the problem of quantum motion 
in a periodic potential was solved already in the 1920's~(see 
e.g.~\cite{kittel}) and gives rise to band spectra and Bloch states. Nowadays, 
the observation of Landau-Zener tunnelling between Bloch waves is at the fronteers 
of research in Bose-Einstein condensates (BEC) in optical 
lattices~\cite{kasevich,cristiani,niu,wu,liu,jona}. 
The external forcing, responsible for Landau-Zener tunnelling, is created by 
either placing the BEC in a gravitational potential~\cite{kasevich} or 
accelerating the optical lattice itself~\cite{cristiani}.

\begin{figure}[t]
\begin{center}\leavevmode 
\includegraphics[width=1.1\linewidth,clip]{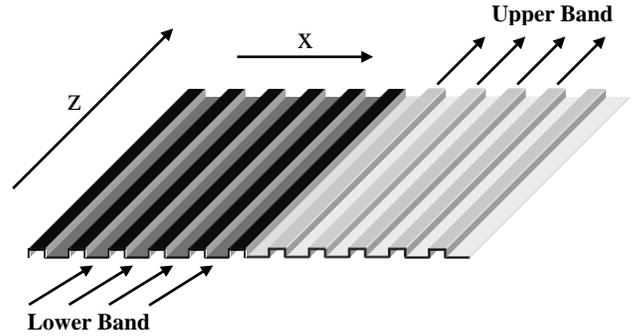}
\vspace{-2cm}
\end{center}
\caption{Schematic picture of the combination of two waveguide arrays of the
same spatial period but with 
different refractive indices (shown by the different grey levels). Choosing 
appropriately [however, see formula (\ref{limits}) 
below] the difference in refractive indices the intensity in the lower band mode is 
almost completely transferred to the upper band mode due to Landau-Zener 
tunnelling.}
\label{array}
\end{figure}

In this Letter we propose a new way of generating Landau-Zener tunnelling.
We consider waveguide arrays~\cite{joseph}, where the periodic potential of
the Schr\"odinger equation is provided by the spatial oscillation of the 
refractive index in the transversal direction. Tunnelling is caused
by combining two waveguide arrays with different refractive indices 
(see Fig.\ref{array}). As we will see, this corresponds to creating a {\it step}
in a periodic potential. 

For arrays of coupled waveguides, the longitudinal direction $z$ (see Fig.\ref{array}),
along which the refractive index is constant, plays the role of ``time" in 
the stationary regime. The refractive index varies only along the transversal
direction $x$, which 
represents space in a $1+1$ (space-``time") dimensional picture. 
Various linear and nonlinear phenomena have been observed in waveguide arrays:
discrete spatial optical solitons~\cite{eisenberg}, diffraction 
management~\cite{silberberg}, excitation of Bloch modes~\cite{mandelik1}, 
generation of multiband optical breathers~\cite{silberberg1} and of 
single band-gap solitons~\cite{mandelik2}, anomalous band-gap 
transmission regimes~\cite{ramaz}. Fast progress in discovering various 
nonlinear effects in waveguide arrays has been possible due to the introduction of 
the {\it tight-binding} approximation~\cite{kivshar1,smerzi,johanson}, which
reduces the nonlinear Schr\"odinger equation to the {\it discrete} nonlinear 
Schr\"odinger equation~\cite{joseph}. 

However, such a reduction eliminates the rich band structure of the periodic medium 
and only a single Bloch band is left. On the contrary, we want to 
maintain the band structure and, hence, study transitions between the bands.
Indeed, as we will see below, the coupling of two waveguides with different
refractive indices introduces a step in the periodic potential and, consequently, 
Landau-Zener tunnelling. In the following we will study only transitions between
the first two bands, denoting them {\it upper} and {\it lower} band.
In particular, we propose to inject light into the left waveguide array
with a given angle in order to populate the lower Bloch band, and retrieve it from 
the right waveguide array. 
Below, we derive analytically the step size bounds inside which most of the intensity 
of the lower band mode is transferred to the upper band mode, creating also a spatial separation between lower and upper band light.
We demonstrate this effect by performing numerical simulations of the
Schr\"odinger equation in a periodic potential with a step.

\begin{figure}[t]
\begin{center}\leavevmode 
\includegraphics[width=0.7\linewidth,clip]{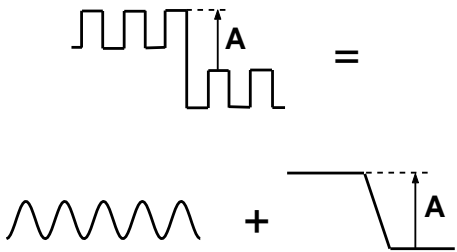}
\end{center}
\caption{Approximation of the periodic potential with a step used in the
Schr\"odinger equation describing the waveguide arrays with different refractive
indices.}
\label{approximation}
\end{figure}

The waveguide array refractive index profile in the transversal direction $x$, which
has a periodic rectangular shape, is approximated by a harmonic potential,
while the step in the refractive index is substituted by a slope $-\alpha$ of
height $A$ (see Fig.~\ref{approximation}).
The functional form of the harmonic potential is selected in order to keep
the ground state at zero energy, as done in the case of BEC~\cite{cristiani}.
Such an approximation allows a simple analytical treatment of the problem. Thus, 
in the linear regime, the adimensionalized Schr\"odinger equation of the optical 
system can be written as follows 
(see e.g. Ref. \cite{kivshar2})
\begin{equation}
i\frac{\partial \Psi}{\partial z}+\frac{1}{2}\frac{\partial^2\Psi}{\partial 
x^2}+\Bigl(V(x)+2w\sin^2x\Bigr)\Psi=0, \label{1}
\end{equation}
where $\Psi$ stands for the complex envelope of the electric field and $w$ is the height of the harmonic potential. Moreover,
\begin{eqnarray} 
&V(x)&=0 \quad \mbox{for}  \quad x<0; \qquad V(x)=-A \quad \mbox{for} \quad 
x>A/\alpha \nonumber \\ &&
\mbox{and} \qquad V(x)=-\alpha x \quad \mbox{for} \quad 0<x<A/\alpha.
\label{step}
\end{eqnarray}

Via the simple transformation $\Psi\rightarrow \Psi \exp[izV(x)]$, wave 
equation (\ref{1}) gets a well known form (see Refs. \cite{liu,wu,jona}):
\begin{equation}
i\frac{\partial \Psi}{\partial z}+\frac{1}{2}\left(\frac{\partial}{\partial x}-
i\alpha z\right)^2\Psi+2w\sin^2(x)\Psi=0 \label{2}
\end{equation}
where $\alpha$ plays the role of acceleration in the Landau-Zener phenomenon.
However, we should keep in mind that acceleration takes place only within the step 
region $0<x<A/\alpha$, unlike the previously considered cases of BEC's, where the 
whole condensate is accelerated (see e.g. Ref. \cite{jona}). Using a two mode 
approximation the wave function $\Psi$ is written as follows
\begin{equation}
\Psi=\left[a(z) e^{iKx}+b(z) e^{i(K-2)x}\right].
\label{twomode}
\end{equation}
With our conventions the zone-boundary mode wavenumber is 1.
It should be mentioned that the two mode approximation works better just in 
the vicinity of zone boundaries ($K\rightarrow 1$), exactly where Landau-Zener 
tunnelling takes place. Following Ref.~\cite{wu}, we substitute Expr.~(\ref{twomode}) into the wave 
Eq.~(\ref{2}). Assuming $K=1$ and removing a common phase dependence in $a(z)$ and 
$b(z)$, we get the Landau-Zener model~\cite{landau,zener} in its original form
\begin{eqnarray}
i\frac{\partial a}{\partial z}&=&-\alpha z a +\frac{w}{2}b \nonumber \\
i\frac{\partial b}{\partial z}&=&\alpha z b +\frac{w}{2}a.
\label{zener}
\end{eqnarray}
Thus, according to Landau-Zener's result, tunnelling from the lower zone-boundary mode
to the upper band takes place with the following rate
\begin{equation}
r=exp\left[-\frac{\pi w^2}{4\alpha}\right].
\label{rate}
\end{equation}
We consider cases in which the acceleration constant $\alpha$ is large. Thus, 
the tunnelling rate is close to one, meaning that almost all light intensity approaching the step is 
transferred to the upper band. This is confirmed by numerical simulations.

\begin{figure}[t]
\begin{center}\leavevmode 
\includegraphics[width=1\linewidth,clip]{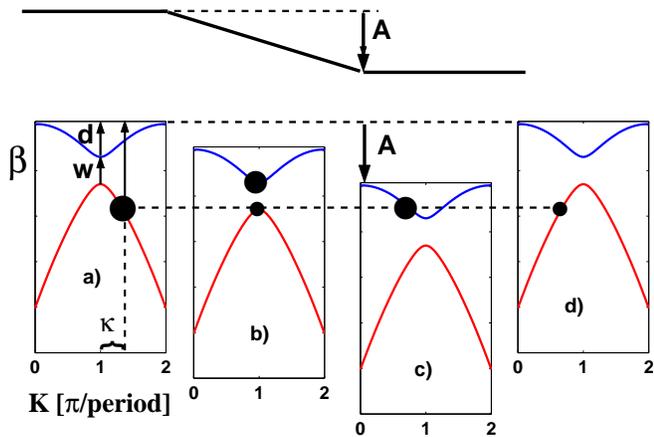}
\end{center}
\vspace{0.1cm}
\caption{Schematic band-gap structure and picture of the Landau-Zener tunnelling process.
$w$ is the gap between the bands (the height of the period potential),
$d$ is the width of the upper band and $\kappa$ is the initial detuning of the lower
band mode wavenumber from zone-boundary. $\beta$ is the dimensionless propagation constant
(see formula~\ref{dispersion}). $A$ is the height of the step.
Initially, light is injected in the left array, populating the lower band
mode(a). Going across the step $\kappa$ decreases, reaching zero as the mode approaches 
the zone-boundary causing Landau-Zener tunnelling (b). After tunnelling, most of the light 
intensity is transferred to the upper band mode, whose wavenumber decreases until
the end of the step is reached. This is the light we observe in the right array (c). 
A smaller light intensity remains in the lower band and is observed in the
left array (d).}
\label{tunnelling}
\end{figure}

Outside the step, where acceleration is absent, one can 
write down  the wave-functions and the dispersion relations in simple 
approximate form (see e.g.~\cite{kittel}). 
\begin{eqnarray}
\Psi_{\pm}&=&\Bigl[\frac{2\kappa\pm \sqrt{w^2+4^2\kappa^2}}{w}e^{-ix}- 
e^{ix}\Bigr]e^{i(\beta_{\pm}z+\kappa x)} 
\label{wavefunction} \\ &&
\beta_{\pm}(\kappa)=\frac{\pm\sqrt{w^2+4\kappa^2}-1-\kappa^2}{2}~,
\label{dispersion}
\end{eqnarray}
where $\kappa=K-1$ is the wavenumber detuning from the zone-boundary and 
the $+$($-$) sign indicates the upper (lower) band. $\beta$ is the dimensionless
propagation constant.
The dispersion relations (\ref{dispersion}) for the two bands are schematically shown in 
Fig.~\ref{tunnelling}. The picture is similar to the one observed in experiments~\cite{mandelik2}. 
Let us remark that at the zone-boundary ($\kappa\rightarrow 0$), the amplitudes of the upper and 
lower band modes are $|\Psi_+|=2\sin x$ and $|\Psi_-|=2\cos x$, respectively.
This means that the light intensity in the lower band is concentrated inbetween  
the waveguides centers, while, in the upper band, intensity is concentrated on 
waveguides centers. This property is a clear experimental indication whether the wave is 
in the lower or upper band.

The experimental setup could be as follows. One should inject a lower band mode with 
non zero but small relative wavenumber $\kappa$. This is accomplished by choosing
for the light beam a direction forming an angle $\theta$ with respect to the
$z$ direction such that $\tan \theta=\kappa$. Hence, the wave front will
move towards the step. An analysis of the dependence of the tunnelling
mechanism on the physical parameters appearing in Fig.~\ref{tunnelling} shows 
that the transition to the upper band mode verifies only if the refraction index step $A$ fulfills the following inequalities
\begin{equation}
\Delta\beta+w<A<\Delta \beta+w+d,
\label{limits}
\end{equation}
where $\Delta\beta=\beta_-(0)-\beta_-(\kappa)$ is the variation of the propagation constant between the initial state and zone-boundary and $d$ is the width of the upper band. 

\begin{figure}[t]
\begin{center}\leavevmode 
\includegraphics[width=1\linewidth,clip]{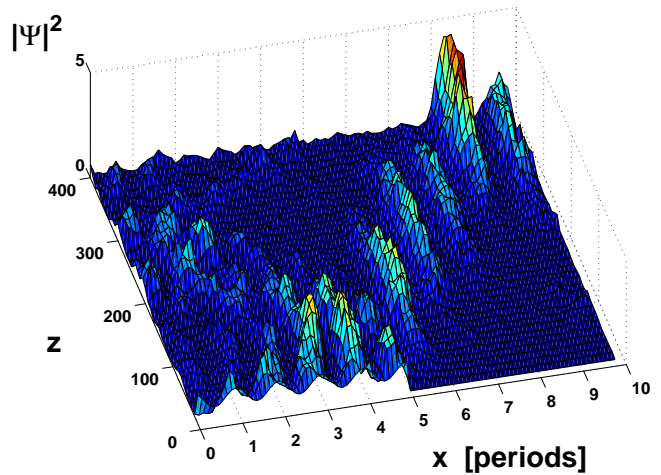}
\end{center}
\caption{Landau-Zener tunnelling from the lower to the upper band. 
Step size $A=0.7$ is taken within the limits of Exp. (\ref{limits}). Step is placed at $x=5$ and lower band mode is injected into the first five periods.}
\label{3D1}
\end{figure}

\begin{figure}[t]
\begin{center}\leavevmode 
\includegraphics[width=1\linewidth,clip]{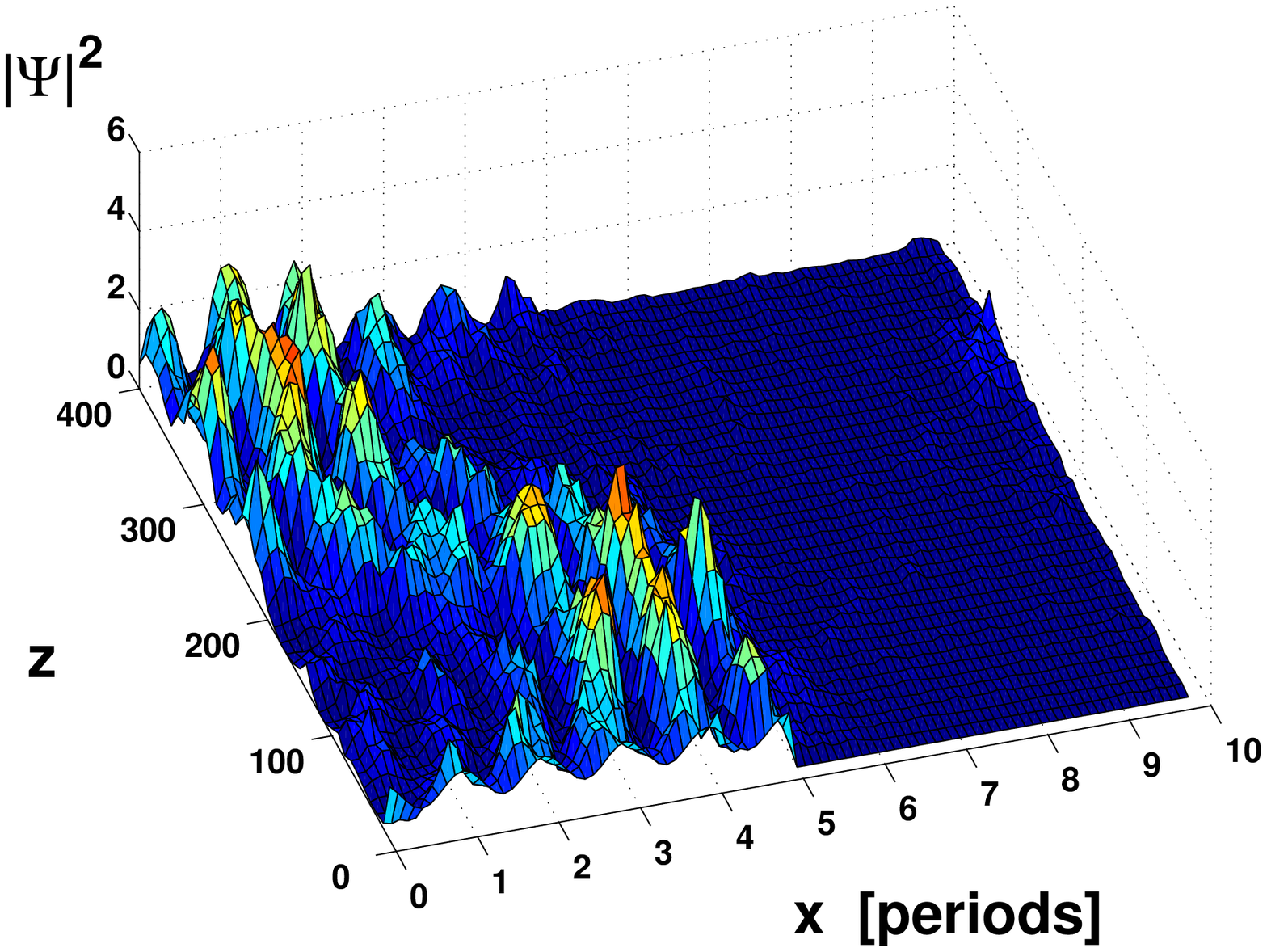}
\includegraphics[width=1\linewidth,clip]{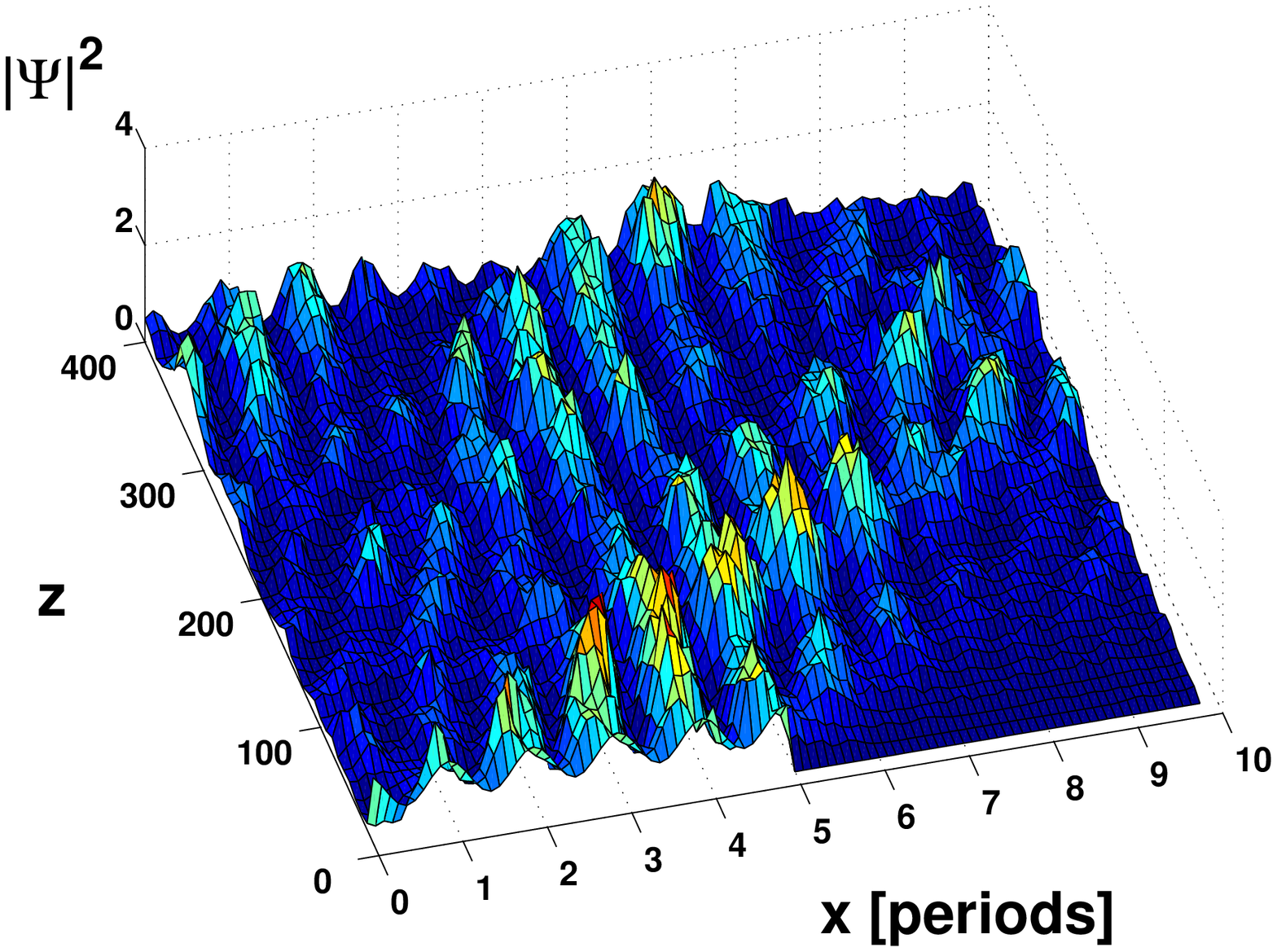}
\end{center}
\caption{
Upper graph: Total reflection from the step. Step size $A=1$ is taken higher 
than the upper bound of Exp. (\ref{limits}).
Lower graph:
Penetration through the step without tunnelling. Step size 
$A=0.07$ is taken within the step values form the inequality (\ref{limits1}).}
\label{3D2}
\end{figure}

Let us try to justify this result by commenting at the same time the
results of some numerical simulations. These are performed by fixing
$w=0.5$, $\kappa=0.2$ and varying the step size $A$. Waveguide centers 
are placed every period, with the first waveguide at half a period from 
the left boundary. The refractive index step is placed in the middle of 
the array. If the refractive index step $A$ is within the above limits (\ref{limits}), 
the lower band mode cannot overcome the step and when it 
reaches the zone-boundary Landau-Zener tunnelling to the upper band occurs. 
Light is partially transmitted to the right in the upper band and reflected
to the left in the lower band. This regime is demonstrated in Fig.~\ref{3D1},
which evidentiates, even visually, the fact that light in the left array is
concentrated inbetween waveguides, while it lies at waveguide centers on
the right array [see also the form of wavefunctions (\ref{wavefunction})].
If the step size is higher than the upper bound of Exp. (\ref{limits}), 
the lower band mode cannot overcome the step and total reflection takes 
place (see the upper graph of Fig.~\ref{3D2}). On the other hand, if 
\begin{equation}
A<\Delta \beta
\label{limits1}
\end{equation}
the lower band mode is able to overcome the step and to penetrate to the right 
side without tunnelling (see the lower graph of Fig.~\ref{3D2}). Indeed, wave 
intensity is now concentrated in-between the waveguides and one can conclude 
that only lower band modes are present in the array.

Finally, if the step is located within the following limits
\begin{equation}
\Delta\beta<A<\Delta\beta+w,
\label{limits2}
\end{equation}
the wave on the left array has no counterpart on the left with the same propagation constant, thus, no stationary penetration of the light through the step is possible, more or less like in the upper Fig.~\ref{3D2}.

Concluding, a novel type of linear optical tunnelling effect is discovered.
This is discussed in connection with waveguide arrays, which have a spatially
oscillating refractive index.
We explain this effect resorting to Landau-Zener model, which is commonly
used for accelerated quantum-mechanical two-level systems.  
In our case, tunnelling between different bands takes place while a wave passes through
a step in the refractive index. In waveguide arrays, Bloch bands in the propagation
constant (light wavenumber) play the role of energy bands in quantum
mechanics. Numerical simulations show that, if  certain limits in the refractive index step
are respected, a spatial separation of light in different bands can be achieved. 
More interesting for applications is, perhaps, the use of this mechanism to
build {\it light wavenumber filters}. Indeed, by appropriately choosing
the physical parameters, it could be possible to shift the central wavenumber of
a light beam and to reduce the spread in both frequency or wavenumber.
The inclusion of a weak nonlinearity does not qualitatively alter the picture, while strongly 
nonlinear cases need a separate treatment. In particular, the scattering 
process of optical gap solitons with the step could be of special 
interest. Moreover, it should be observed that the effect discussed in this 
Letter is generic for systems with periodic potential and that
applications in different fields could be found. For 
instance, the analysis of a similar process in Bose-Einstein condensates 
is in progress. 

We thank E. Arimondo, M. Jona-Lasinio, Yu. Kivshar and D. Mandelik for useful suggestions. This work is funded by the contract COFIN03 of the Italian MIUR {\it Order and chaos in nonlinear extended systems}. R.K. acknowledges hospitality from Dipartimento di Energetica "S. Stecco" (Florence University, Italy), the Abdus Salam International Center for Theoretical Physics (Trieste, Italy) and financial support in the frame of CNR-NATO Senior fellowship award No 217.35 S.

\end{document}